\documentclass[a4paper,11pt]{article}
\usepackage{amssymb}
\usepackage{graphicx}
\usepackage{epsfig}
\usepackage{subeqn}
\usepackage{subfigure}
\usepackage{supertabular}

\newcommand{\beq}{\begin{equation}}
\newcommand{\eeq}{\end{equation}}
\newcommand{\beqa}{\begin{eqnarray}}
\newcommand{\eqa}{\end{eqnarray}}

\title{Magnetic plateaus in the 1D antiferromagnetic spin-3/2 and spin-2 Ising chains with single-ion
anisotropy}

\author{\centerline{Ekrem Ayd\i ner, Cenk
Aky\"{u}z, Meltem G\"{o}n\"{u}lol and Hamza Polat}\\
\textit{Theoretical and Computational Physics Research Group,
Department of Physics}\\
\textit{Dokuz Eyl\"{u}l University, Tr-35160 \.{I}zmir, Turkey}\\
\textit{E-mail: ekrem.aydiner@deu.edu.tr}}

\begin{document}
\maketitle

\begin{abstract}
In this study, we have employed the classical transfer matrix
technique to investigate the magnetization plateaus, phase
diagrams and other thermodynamical properties of the
one-dimensional antiferromagnetic spin-3/2 and spin-2 Ising chains
with single-ion anisotropy in the presence of an external magnetic
field at very low temperature. We have showed that single
ion-anisotropy is one of the indispensable ingredients for an
energy gap which leads to magnetic plateau mechanism in
one-dimensional antiferromagnetic Ising spin chains. Other
thermodynamical predictions seem to be provide this argument.
\end{abstract}

\vspace*{25pt}


\newpage
\section{Introduction}
Low-dimensional (one-dimensional (1D) and quasi-one dimensional)
gapped spin systems such as spin-Peierls, Haldane and spin ladder
systems have drawn attention from both theorists and
experimentalists in the literature after the prediction by Haldane
\cite{haldane} that a 1D Heisenberg antiferromagnet should have an
energy gap between the singlet ground state and the first excited
triplet states in the case of an integer spin quantum number $S$,
while the energy levels are gapless in the case  of a half-odd
integer values of $S$.

The most fascinating characteristic of these systems is to show
magnetic plateaus i.e. topological quantization of the
magnetization at the ground state of the system due to magnetic
excitations. A general condition for the quantization of the
magnetization in low-dimensional magnetic systems is derived from
the Lieb-Schultz-Mattis theorem \cite{lieb}. Oshikawa, Yamanaka
and Affleck (OYA) \cite{oshikawa} discussed this plateau problem
and derived a condition $p(S-m)=integer$, necessary for the
appearance of the plateau in the magnetization curve of
one-dimensional spin system, where $S$ is the magnitude of spin,
$m$ is the magnetization per site and $p$ is the spatial period of
the ground state, respectively. This condition emphasize that the
$2S+1$ magnetization plateaus (contained the saturated
magnetization $m=S$) can appear when the magnetic field increases
from zero to its saturation value $h_{s}$, however, it does not
directly prove its existence \cite{strecka}.

In many theoretical studies, it has been found that
low-dimensional gapped spin systems have magnetic plateaus at low
temperatures or ground state as providing OYA criterion. For
example, Chen et al. \cite{chen} employed the classical Monte
Carlo technique to investigate the magnetization plateaus of 1D
classic spin-$1$ and spin-$3/2$ AF Ising chain with a single-ion
anisotropy under the external field at low temperatures, and they
showed that the systems have $2S+1$ magnetic plateaus. Tonegawa et
al. \cite{tonegawa} observed the plateau at $m=0.0, 0.5$, and
$1.0$ in the ground state of spin-$1$ AF Heisenberg chain with
bond alternating and uniaxial single-ion anisotropy. Sato and
Kindo obtained a magnetic plateau at $m=0.0, 0.5$, and $1.0$ for
spin-$1$ bond alternating Heisenberg chain by using
renormalization group technique \cite{satokindo}. Hida observed
that an $S=\frac{1}{2}$ antiferromagnetic chain with period-3
exchange coupling shows a plateau in the magnetization curve at
magnetization per site $m=\frac{1}{6}$ \cite{hida}. Totsuka found
a plateau at $m=\frac{1}{4}$ for the familiar $S=\frac{1}{2}$
antiferromagnetic Heisenberg chain with the next-nearest-neighbor
and alternating nearest-neighbour interactions using the
bosonization technique \cite{totsuka1}. Sakai and Takahashi found
that a magnetization plateau appears at $m=\frac{1}{2}$ for the
$S=\frac{3}{2}$ antiferromagnetic Heisenberg chain by the exact
diagonalization of finite clusters and finite-size scaling
analysis \cite{sakaitakahashi}. Yamamoto \textit{et al.} obtained
half-saturated plateau of $m=\frac{1}{4}$ in the $S=\frac{1}{2}$
XXZ chain in $XY$ phase exposed to a weak period-4 field
\cite{yamamoto}. Oshikawa \textit{et al.} numerically obtained the
magnetic plateau $m=\frac{1}{2}$ in the $S=\frac{3}{2}$
antiferromagnetic Heisenberg chain with single ion anisotropy
$D\geq2$ \cite{oshikawa}. Ayd\i ner and Aky\"{u}z obtained
magnetic plateaus for spin-1 using classical transfer matrix
technique \cite{aydiner1}. Okamoto also found out the
magnetization plateau in alternating spin chains \cite{okamoto}.

On the other hand, magnetic plateaus have been predicted not only
in theoretical study but also have been observed in experimental
studies. For example, Narumi et al. \cite{narumi1,narumi2}
observed the magnetic plateaus in the magnetization curve for both
[Ni$_{2}$(Medpt)$_{2}$($\mu$-ox)(H$_{2}$O)$_{2}$](ClO$_{4}$)$_{2}$2H$_{2}$O
and [Ni$_{2}$(Medpt)$_{2}$($\mu$-ox)($\mu$-N$_{3}$)](ClO$_{4}$)0.5
H$_{2}$O. Goto et al. \cite{goto} reported the existence of the
magnetization plateau at $0.25$ in spin-$1$ 3,3',5,5'-tetrakis
(N-tert-butylaminxyl) biphenyl (BIP-TENO).

In the majority of plateau mechanisms which have been proposed up
to now the purely quantum phenomena play crucial role. The concept
of magnetic quasi-particles and strong quantum fluctuation are
regarded to be first importance of these processes. Particularly,
for a number of systems it was shown that the plateau at $m\neq0$
are caused by the presence of the spin gap in the spectrum of
magnetic excitation in the external magnetic field \cite{ohanyan}.
Another mechanism lies in so-called crystallization of the
magnetic particles \cite{totsuka2,momoi}. Meanwhile, even though
the phenomenon of magnetization plateaus are often regarded to
have purely quantum origin, it is obviously not clear since
one-dimensional \cite{chen,aydiner1,ohanyan,aydiner2} and two
dimensional \cite{kubo,honecker} quasi-classical spin models which
exhibit a magnetization plateaus at zero temperature or near the
ground state. In fact, these studies show to us that Heisenberg
and Ising spins leads to the qualitatively same structures of the
magnetization profiles, particularly to the formation of
magnetization plateaus. According to our knowledge, magnetic
plateaus can appear depend on dimerization, frustration,
single-ion anisotropy or periodic field for one-dimensional AF
spin chain independent of whether the classical or not quantum
models. Therefore, we say that one of these ingredients is
required for a plateau mechanism.

In this study, we shall investigate the one-dimensional
antiferromagnetic spin-3/2 and spin-2 Ising chains with single-ion
anisotropy to obtain magnetic plateaus and other thermodynamical
quantities taking into account the analogy of the magnetization
profiles between Heisenberg and Ising spin systems. The approach
is based on very simple idea to use the Ising spin instead of the
Heisenberg operators and to develop the classical transfer matrix
technique. It is worth to emphasize that in contrast to the
majority of existing approaches to the problem of magnetization
plateau, this technique is entirely based on analytical
calculations and allows to obtain the magnetization profiles for
arbitrary finite temperature and arbitrary values of $J$, $D$ and
$h$. On the other hand, we must remark that the transfer matrix
approach to the 1D classical system is rather simple as compared
with other techniques, however, the implementation of this
technique leads to remarkable simplicity and clarity and produces
considerably exact results for 1D infinite spin system.

The organization of this paper is as follows. In the next section,
we present classical Hamiltonian for spin-$S$ system with
single-ion anisotropy and briefly introduce transfer matrix
technique. In section 3 and 4 we give numerical results for
spin-3/2 and spin-2, respectively. The last section gives our
conclusions.

\section{Theoretical Approach}
In point of classical view, Hamiltonian for the 1D spin-$S$ system
with single-ion anisotropy can be written as follows:
\begin{equation}\label{this}
H=J\sum\limits_{i=1}^{N}S^{z}_{i}S^{z}_{i+1}+h\sum\limits_{i=1}^{N}S^{z}_{i}+
D\sum\limits_{i=1}^{N}{(S^{z}_{i})}^2,
\end{equation}
where $J$ denotes the exchange coupling of antiferromagnetic type
(i.e $J>0$), $D$ is the single-ion anisotropy, $h$ is the external
field, and \emph{N} is the total number of site in the system.

Herein, in order to investigate the magnetic and thermal
properties of 1D AF spin-$3/2$ and spin-$2$ Ising chain, we are
interested in the three thermodynamical expressions (i)
ferromagnetic order $m$, (ii) the specific heat $C$, and (iii) the
magnetic susceptibility $\chi$. These quantities are calculated in
terms of free energy as
\begin{subequations}
\begin{equation}
m=-\frac{\partial{f(h,D,T)}}{\partial{h}},
\end{equation}
\begin{equation}
C=-T\frac{\partial^2{f(h,D,T)}}{\partial{T^2}},
\end{equation}
\begin{equation}
\chi=-\frac{\partial^2{f(h,D,T)}}{\partial{h^2}}.
\end{equation}
\end{subequations}
Then the free energy for the system can be found exactly in terms
of the partition function with the help of the transfer matrix
technique. Hence, all thermodynamic functions in Eq. (2) are
respectively calculated.

The free energy per spin of this system is given by
\begin{equation}
f(h,D,T)=-kT\lim\limits_{N\rightarrow\infty}\frac{1}{N}\ln\mathcal{Z}.
\end{equation}
with
\begin{equation}
\mathcal{Z}=Tr \textit{V}^N.
\end{equation}
where $V$ represents transfer matrix of the spin system, which we
give below. It is known that a conventional way to calculate the
free energy is to write the partition function (4) in terms of
eigenvalues of the transfer matrix $V$ as follows:
\begin{equation}
\mathcal{Z}=Tr\textit{V}^N
=\lambda_{1}^N+\lambda_{2}^N+\lambda_{3}^N+... .
\end{equation}
According to the standard assumptions of this technique, partition
function (5) is represented as,
\begin{equation}
\mathcal{Z}=Tr{V}^N
=\lambda_{1}^N[1+({\frac{\lambda_{2}}{\lambda_{1}}})^N+
({\frac{\lambda_{3}}{\lambda_{1}}})^N+...],
\end{equation}
where $\lambda_{1}$, $\lambda_{2}$, $\lambda_{3}$,... which are
sorted from the biggest to smallest.

It is clearly seen that the second and the third terms on the RHS
of Eq. (6) go to zero as $N\rightarrow{\infty}$ due to
${\mid{\frac{\lambda_{2}}{\lambda_{1}}}\mid}<1$,
${\mid{\frac{\lambda_{3}}{\lambda_{1}}}\mid}<1$, and so on. Hence
the free energy Eq. (3) is practically reduced to
\begin{equation}
f(h,D,T)=-kT\ln{\lambda_1}.
\end{equation}
This result implied that the magnetic and thermodynamical
quantities in Eq. (2) for the present system can be evaluated
depend on the biggest eigenvalue $\lambda_1$.

For spin-3/2, $S^{z}_{i}$ takes on $\pm\frac{3}{2}$, and
$\pm\frac{1}{2}$ values, on the other hand, for spin-2,
$S^{z}_{i}$ takes on $\pm{2}$, $\pm{1}$, and ${0}$ values.
 Considering the fact that, for example, we can write the
transfer matrix $V$ for spin-3/2 as

\begin{equation}
V=\left(%
\begin{array}{cccc}
  e^{(9\alpha+9\eta+3\xi)} & e^{(3\alpha+ 5\eta+2\xi)}& e^{(-3\alpha+5\eta+\xi)}&
  e^{(-9\alpha+9\eta)} \\ e^{(3\alpha+5\eta+2\xi)} & e^{(\alpha+\eta+\xi)}&
  e^{(-\alpha+\eta)} & e^{(-3\alpha+5\eta-\xi)} \\ e^{(-3\alpha+5\eta+\xi)} &
  e^{(-\alpha+\eta)} & e^{(\alpha+\eta-\xi)} & e^{(3\alpha+5\eta-2\xi)} \\
  e^{(-9\alpha+9\eta)} & e^{(-3\alpha+5\eta-\xi)} & e^{(3\alpha+5\eta-2\xi)} &
  e^{(9\alpha+9\eta-3\xi)} \\
\end{array}%
\right)
\end{equation}
where $\beta=1/kT$, $\alpha=\frac{1}{4}\beta{J}$,
$\eta=\frac{1}{4}\beta{D}$, and $\xi=\frac{1}{2}\beta{h}$ where
$k$ is the Boltzmann constant and $T$ is the temperature. The
transfer matrix $V$ for spin-2 can be constructed in a similar
way.

\section{Numerical Results for AF spin-3/2 Ising chain}

The numerical results for spin-3/2 are as follow: In Fig. 1(a) and
(b), the magnetization $m$ were respectively plotted as a function
of external field $h$ for fixed values of $D=0.5$, and $1.5$ at
$T=0.01$ (unit by $J$). It is seen from Fig. 1(a) and (b) that
$2S+1=4$ plateaus appear, located at $m=0.0, 0.5, 1.0$, and $1.5$,
which satisfies the OYA's criterion for all values of positive
single-ion anisotropy $D$ at the low temperature near the ground
state under the external field $h$. Appearance of the magnetic
plateaus indicate that the 1D AF spin-3/2 Ising chain has a gap
mechanism owing to single ion-anisotropy. Previous theoretical
studies \cite{chen,aydiner1,aydiner2} show that single
ion-anisotropy play a crucial role to appear magnetic plateaus for
classical Ising chains. Our result also confirms previous
theoretical predictions \cite{chen}.
%figure 1
\begin{figure}
\begin{center}
\subfigure[\hspace{0.4cm}]{\label{fig:sub:a}
\includegraphics[width=6cm,height=5.5cm,angle=0]{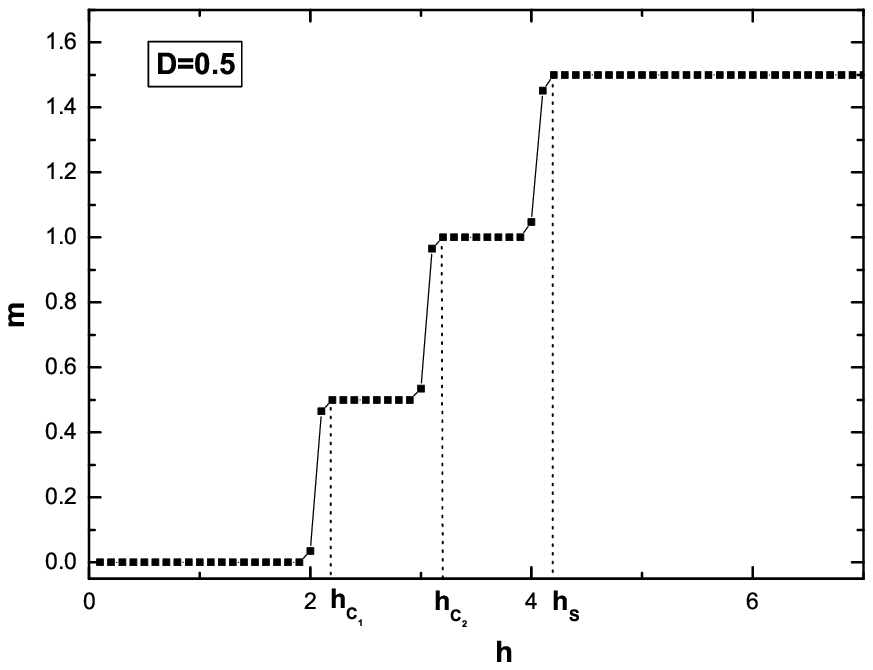}}
\hspace{0.4cm} \subfigure[\hspace{0.4cm}]{\label{fig:sub:b}
\includegraphics[width=6cm,height=5.5cm,angle=0]{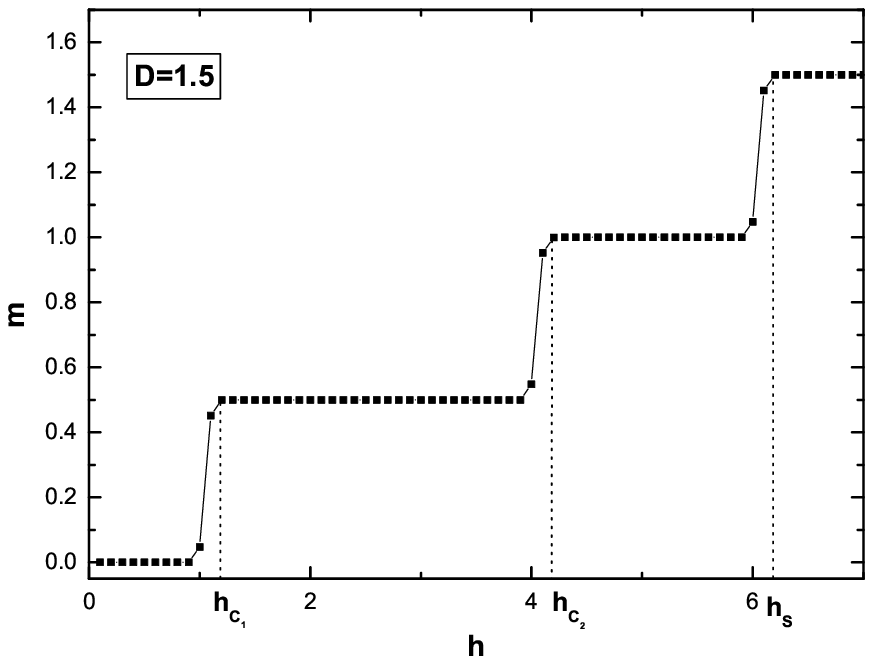}}
\caption{The magnetization $m$ as a function of magnetic field $h$
(a) for $D=0.5$, (b) for $D=1.5$. Here $h_{c_1}$ initial critical,
$h_{c_2}$ second critical, and $h_s$ saturated fields,
respectively. ($T=0.01$, unit by $J$).} \label{fig:sub:Fig1a-1b}
\end{center}
\end{figure}

The transitions between magnetic plateaus occur at critical field
values. The nonzero magnetization at $m=0.5$ begins from a finite
external field which called as initial critical field $h_{c_1}$.
Second plateau at $m=1.0$ appears at second critical field value
$h_{c_2}$. The last plateau corresponds to the value of saturation
magnetization of the system, that we called it as saturated field
$h_s$. On the other hand, we have no observed a magnetization
plateau for $D=0.0$. This observation indicate that single
ion-anisotropy is one of the necessary ingredient to plateau
mechanism for the present system. In fact, $D$ values change width
of plateaus as seen in Fig. 1(a) and (b). In order to see this
effect, magnetic phase diagram $h-D$ was plotted in Fig. 2(a).
This phase diagram displays the behavior of the system depend on
single ion-anisotropy near the ground temperature i.e., $T=0.01$.
However, this phase picture does not clearly show the effect of
exchange interaction $J$.
% figure 2
\begin{figure}
\begin{center}
\subfigure[\hspace{0.4cm}]{\label{fig:sub:a}
\includegraphics[width=6cm,height=5.5cm,angle=0]{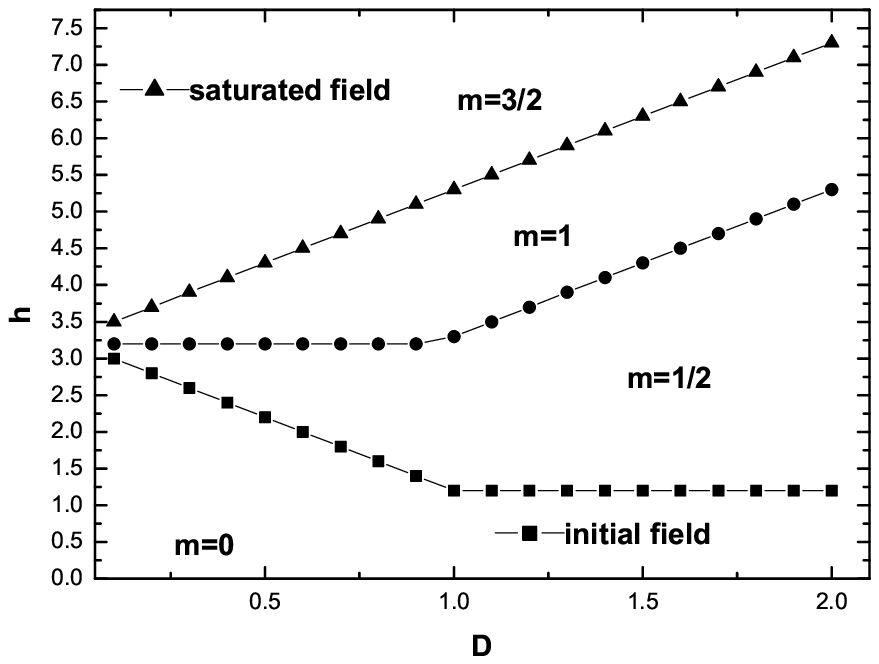}}
\hspace{0.4cm} \subfigure[\hspace{0.4cm}]{\label{fig:sub:b}
\includegraphics[width=6cm,height=5.5cm,angle=0]{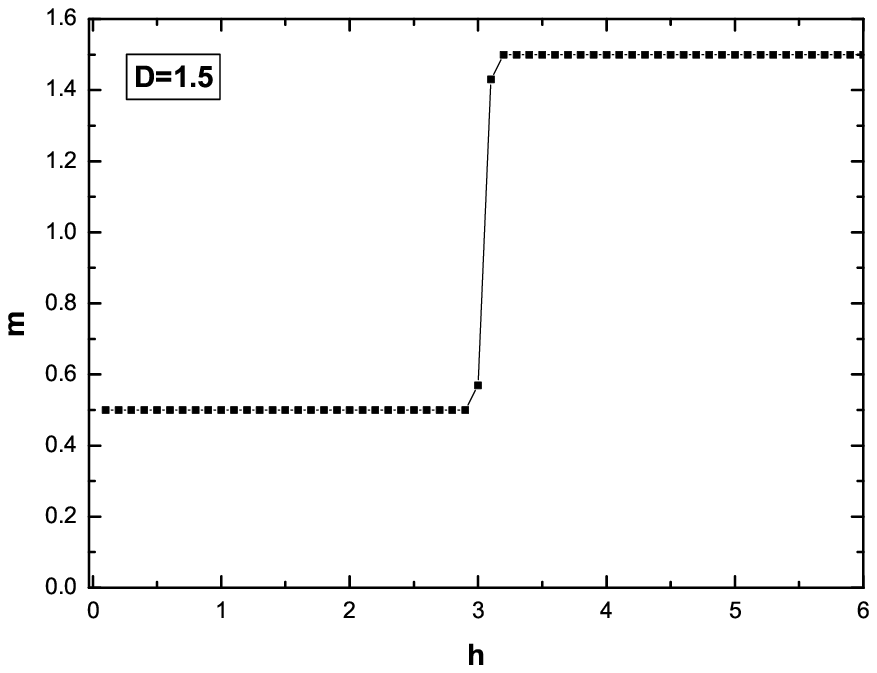}}
\caption{(a) The magnetization phase diagram of the ground state
of antiferromagnetic Ising chain with single-ion anisotropy under
finite magnetic field. The square-dot line, the circular-dot line,
and the upper triangle-dot line represent the initial critical
field, second critical field, and the saturated field,
respectively. (b) In the limit of $J=0$, the magnetization $m$ as
a function of external field $h$ at $D=1.5$ ($T=0.01$, unit by
$J$).} \label{fig:sub:Fig2a-2b}
\end{center}
\end{figure}

Three lines in Fig. 2(a) correspond to critical field values which
are responsible of plateau transition. Four plateau areas i.e.,
$m=0.0$, $0.5$, $1.0$, and $1.5$ are divided by the initial
critical field $h_{c_1}$, second critical field $h_{c_2}$ and the
saturated field $h_s$ lines for values of $D>0.0$ as shown in
Fig.\,2(a). Areas between these lines give information how width
of the plateaus can change depend on single ion-anisotropy. The
longitudinal coordinate of the square-dot is the beginning point
of the field for appearance of the plateau $m=0.5$ which
corresponds to the first critical field $h_{c_1}$ values, the
plateau $m=1.0$ is denoted by the circular-dot which corresponds
to the second critical field $h_{c_2}$ values, and its ending
point is signed by the triangle-dot which correspond to the
saturated field $h_{s}$ values. It is clearly seen that the
initial field line and second critical field line have a critical
behavior at the $D=1.0$ while the saturated field increasing
monotonously with the rise of $D$. The first critical field
decreases in the regime $0.0<D<1.0$ with the increase of $D$,
however, it lies longitudinal for $D>1.0$. On the other hand, the
second critical field increasing monotonously with the rise of $D$
for $D>1.0$, while it lies longitudinal for $D<1.0$. As seen in
Fig. 2(a), value of $D=1.0$ behaves as a critical point for both
first and second critical field values. Reason of this behavior is
chosen as $J=1$. The magnetization phase diagram in Fig.\,2(a) is
compatible with the Monte Carlo prediction \cite{chen}.
% figure 3
\begin{figure}
\begin{center}
\subfigure[\hspace{0.5cm}]{\label{fig:sub:a}
\includegraphics[width=6cm,height=5.5cm,angle=0]{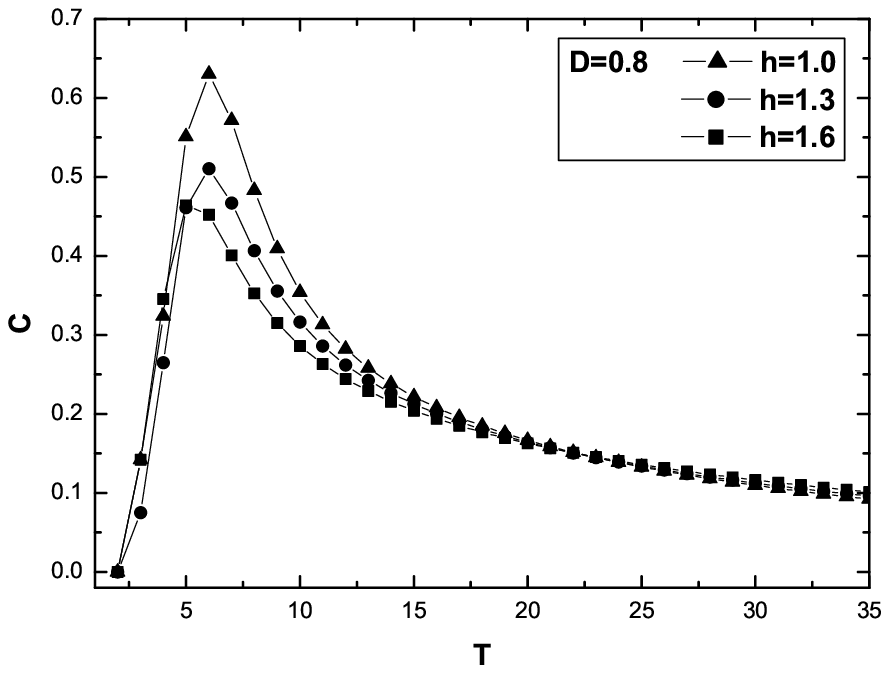}}
\hspace{0.4cm} \subfigure[\hspace{0.5cm}]{\label{fig:sub:b}
\includegraphics[width=6cm,height=5.5cm,angle=0]{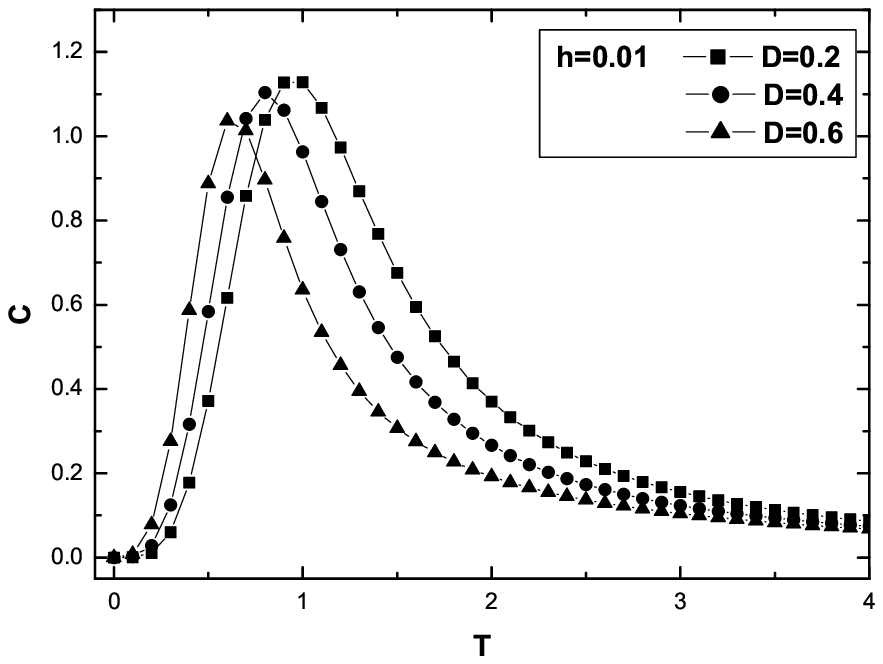}}
\caption{The specific heat $C$ as a function of temperature T
(unit by $J$): (a) $h=1.0, 1.3, 1.6$ at fixed value of $D=0.8$ (b)
$D=0.2, 0.4, 0.6$ at fixed value of $h=0.01$.}
\label{fig:sub:Fig3a-3b}
\end{center}
\end{figure}

The width of the plateau $m=0.5$ increases with $D$ increasing for
the present system, whereas it remains in Heisenberg chain with
single-ion anisotropy $D>0.93$. Furthermore, the width of the
plateaus $m=0.0$ and $m=1.0$ are independent of the single-ion
anisotropy $D$, which is similar to that of $m=0.5$ in
one-dimensional $S=1$ antiferromagnetic Ising chain with
single-ion anisotropy \cite{chen,aydiner1}.

Though the magnetic phase diagram in Fig. 2(a) of the Hamiltonian
(1) gives information about depending single ion-anisotropy,
however, it does not display how the spin exchange strength $J$
contribute to plateau mechanism. In order to understand effect of
$J$ to the plateau, we have studied in the limit of
$\frac{D}{J}\rightarrow{\infty}$. For the simplicity we have
chosen an extreme values of $J=0.0$ and $D=1.5$ and magnetization
were plotted versus external field for these values in Fig.\,2(b).
It is seen that the only intermediate plateau $m=0.5$ expands to
depress the range of the other plateaus $m=0.0$ and $m=1.0$ in
this limit. This result also indicate that spin exchange parameter
$J$ plays a significant role on the plateau mechanism as well as
single ion-anisotropy $D$. At the extreme case of $J=0.0$, we can
easily understand that it is equal to single-particle system with
two plateaus $m=0.5$ and $1.5$ in the ground magnetization.

In order to obtain the behavior of the specific heat $C$ depending
on the external field $h$ and the single-ion anisotropy $D$ at low
temperatures, they are plotted as a function of temperature $T$,
for the fixed value of the single-ion anisotropy, $D=0.8$, at
various values of the external field $h=1.0, 1.3$, and $1.6$ in
Fig.\,3(a) and for the fixed value of $h=0.01$ at various values
of $D=0.2, 0.4$, and $0.6$ in Fig.\,3(b), respectively. The
specific heat curves in Fig.\,3(a) and Fig.\,3(b) have a
relatively board maximum like the Schottky peak near $T=8.0$ and
$T=1.0$ respectively. As can be seen from Fig.\,3(a), the height
of the specific heat peak for the fixed value of $D$ decreases and
moves towards zero temperature when the external field $h$
increases. On the other hand, the height of the specific heat peak
for the fixed value of $h$ decreases and moves towards zero
temperature when the value of single-ion anisotropy increases in
Fig.\,3(b). We note that the Schottky-like round hump in the
specific heat probably reflects the fact that the energy of the
system depends on the external field $h$, and the single-ion
anisotropy $D$. However, they do not indicate that a second-order
phase transition occurs in the one-dimensional system. All of them
may be thought as a Schottky-like peak resulting from the
antiferromagnetic short-range order.

% figure 4
\begin{figure}
\begin{center}
\subfigure[\hspace{0.5cm}]{\label{fig:sub:a}
\includegraphics[width=6cm,height=5.5cm,angle=0]{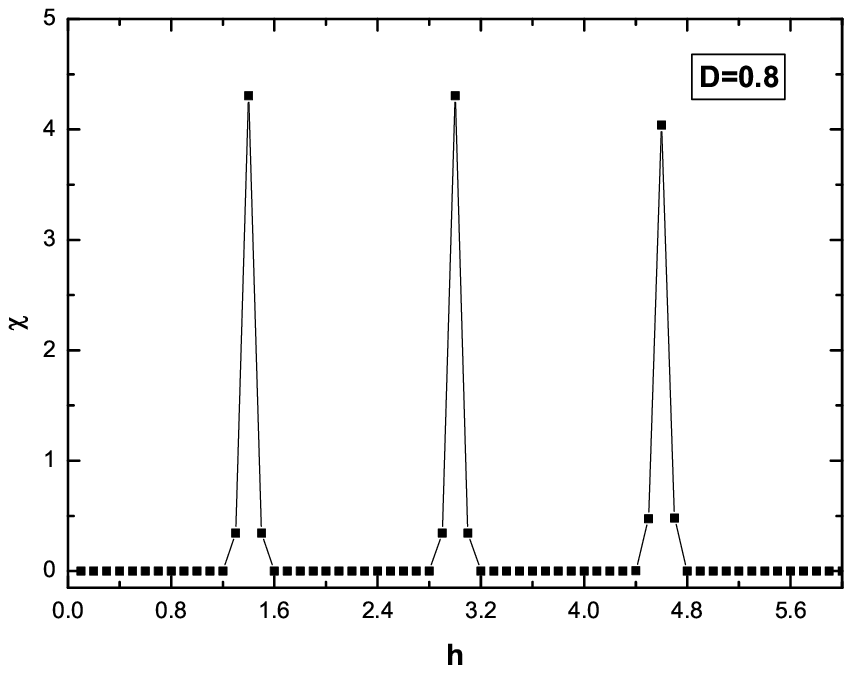}}
\hspace{0.4cm} \subfigure[\hspace{0.5cm}]{\label{fig:sub:b}
\includegraphics[width=6cm,height=5.5cm,angle=0]{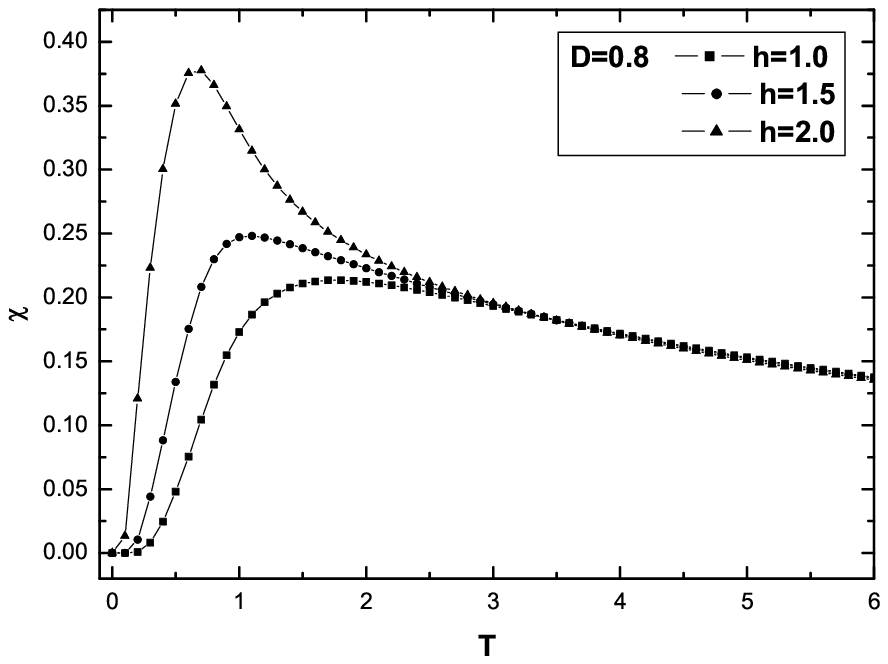}}
\caption{(a) Susceptibility $\chi$ as a function of magnetic field
h at fixed value of $D=0.8$. (b) Susceptibility as a function of
temperature T for various value of $h$ at $D=0.8$(unit by $J$).}
\label{fig:sub:Fig4a-4b}
\end{center}
\end{figure}

Another way of the confirmation of multi-plateau magnetization
curves by numerical calculation in the 1D AF Ising chain is to
examine the field and temperature dependence of susceptibility
$\chi$ in Fig.\,4(a) and (b), respectively. It is expected that
the temperature dependence of the susceptibility gives information
that whether the low-dimensional system has an energy gap or not,
while field dependence of it verify formation of plateaus.
Therefore, the susceptibility is plotted in Fig.\,4(a) as a
function of external field $h$. It can be seen that there are
three peaks in the susceptibility for $D=0.8$ and $T=0.01$. The
first peak occurs at the initial critical field value, the second
peak occurs at the second critical field value and the third peak
occurs at the saturated field value for the fixed value of $D$.
Each peak indicates the transitions from one magnetic plateau to
another one. Furthermore, we have also plotted the magnetic
susceptibility as a function of temperature for fixed value of
$D=0.8$ and for three selected external fields $h=1.0, 1.5$, and
$2.0$ in Fig.\,4(b). The curve for $h=1.0$ marked by
triangular-dot line show a relatively sharp peak at low
temperature, while the other curves for $h=1.5$, and $2.0$ have a
broad maximum which are denoted by circular-dot and square-dot
line, respectively. We roughly say that the susceptibility curves
decay exponentially with the increasing temperature by the
relation $\chi\left(T\right)\approx\exp\left(-E_{g}/k_{B}T
\right)$. The peaks or round hump behavior in the magnetic
susceptibility curves in Fig.\,4(b) may be interpreted as evidence
the formation of the short-range correlations within the chains.
In that case, these correlations may play a role for gap mechanism
just as in Heisenberg AF systems. This characteristic behavior in
the susceptibility clearly compatible with the experimental
results for spin-$3/2$ compound Cs$_{2}$CrCl$_{5}$.4H$_{2}$O
\cite{chatterjee}.

\section{Results for AF spin-2 Ising chain}

We start our discussion once again with investigating the
properties of one-dimensional AF spin-$2$ Ising chain with
single-ion anisotropy. Now, we consider the above Hamiltonian (1)
for the case of spin-$2$. In the case of spin-3/2 the physical
quantities such as the magnetization, the magnetization phase
diagram, the specific heat, and the magnetic susceptibility were
evaluated numerically to keep in this system of spin-2.

We have plotted the magnetization $m$ as a function of external
field $h$ for only fixed values of $D=1.5$ and $T=0.01$ (unit by
$J$) as shown in Fig. 5(a). The number of the plateaus obey to OYA
criterion for spin-2 chain in the case of the spin-$3/2$ chain.
The plateaus appear at $m=0.0$, $0.5$, $1.0$, $1.5$, and $2.0$ for
all values of positive single-ion anisotropy $D$. It is seen from
Fig. 5(a) that there are three critical fields as $h_{c_1}$,
$h_{c_2}$ and $h_{c_3}$ in the spin-$2$ chain. On the other hand,
we have plotted the magnetization $m$ as a function of the
external field $h$ for only fixed values of $D=1.5$ and  for the
extreme case of $J=0.0$ in Fig. 5(b) to see behavior of the
present system in the limit of $\frac{D}{J}\rightarrow{\infty}$.
In this limit, the spin-2 Ising chain has the three plateaus at
$m=0.0$, $1.0$, and $2.0$. At the extreme case of $J=0.0$, one can
easily suggest that it is equal to two-particle system with three
plateaus at $m=0.0$, $1.0$, and $2.0$ in the ground magnetization.

The single ion-anisotropy dependence of the magnetization plateaus
of 1D AF spin-2 Ising chain near the ground state were presented
in $h-D$ space in Fig. 6. The characteristic behavior of the phase
diagram of the spin-$2$ chain is similar to that of the spin-$3/2$
chain. The saturated field increasing monotonously with the rise
of $D$, however, all critical field values i.e., $h_{c_1}$,
$h_{c_2}$, and $h_{c_3}$ have a critical behavior at the $D=1.0$
due to choosing of $J=1.0$. The first and second critical fields
decrease in the regime $0.0<D<1.0$ with increase of $D$, but they
increasing monotonously with the rise of $D$. The third critical
field increase slowly with rise of $D<1.0$, however, it increase
monotonously with the rise of $D$. On the other hand, the width of
the plateau $m=1.0$ increases with $D$ increasing for the present
system, however, the width of the plateaus $m=0.5$ and $m=1.0$
remain constant for values of $D>1.0$, while they increase with
$D$ increasing up to $D=1.0$ as seen from Fig. 6.

% figure 5
\begin{figure}
\begin{center}
\subfigure[\hspace{0.5cm}]{\label{fig:sub:a}
\includegraphics[width=6cm,height=5.5cm,angle=0]{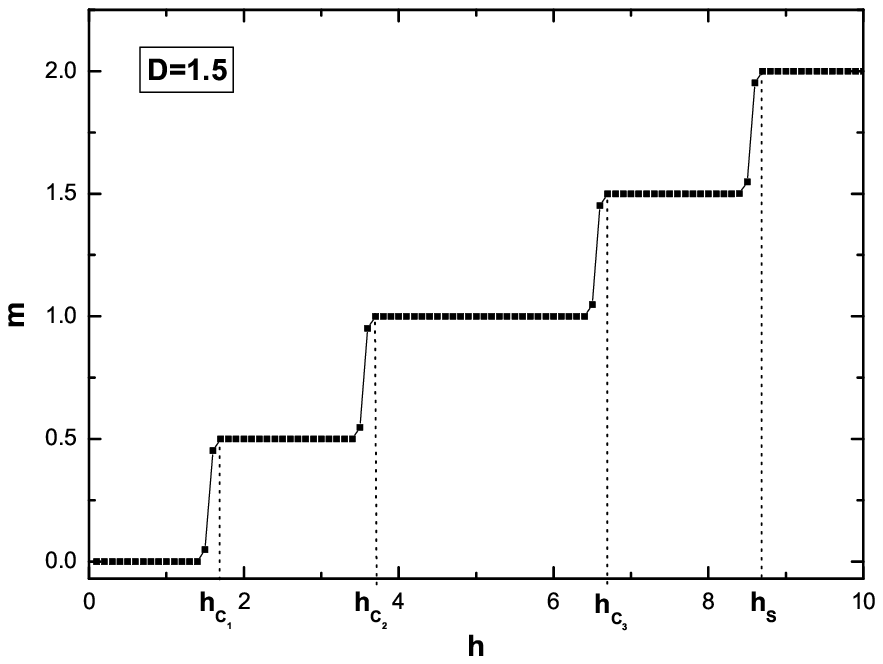}}
\hspace{0.4cm} \subfigure[\hspace{0.5cm}]{\label{fig:sub:b}
\includegraphics[width=6cm,height=5.5cm,angle=0]{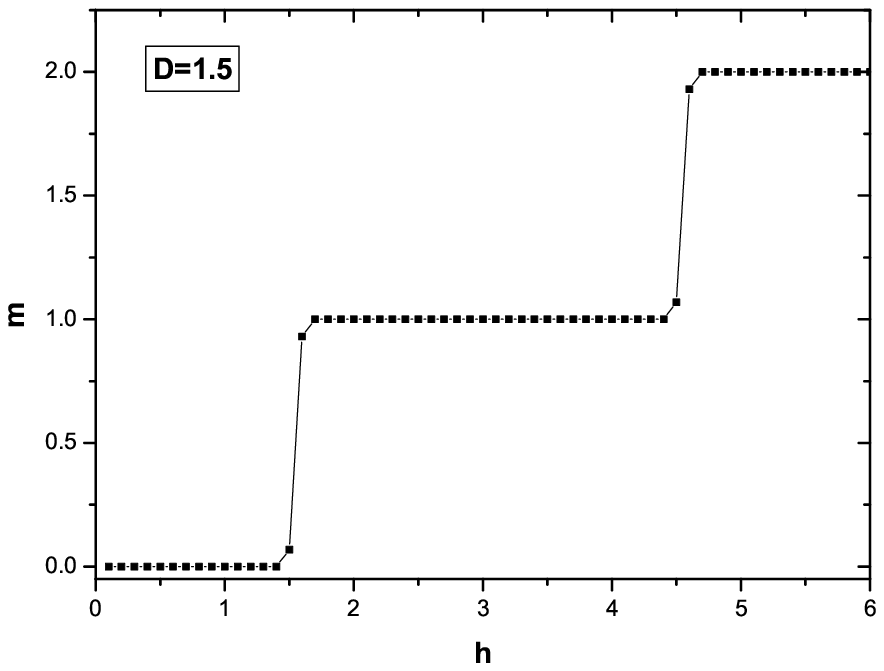}}
\caption{(a) The magnetization $m$ as a function of magnetic field
$h$ for $D=1.5$. Here $h_{c_1}$ the initial critical field,
$h_{c_2}$ the second critical field, $h_{c_3}$ the third critical
field, and $h_s$ saturated field, respectively ($T=0.01$, unit by
$J$) (b) In the extreme case of $J=0$, the magnetization $m$ as a
function of magnetic field $h$ for $D=1.5$ ($T=0.01$). }
\label{fig:sub:Fig5a-5b}
\end{center}
\end{figure}

% figure 6
\begin{figure}
\begin{center}
\includegraphics[width=7.5cm,height=6.5cm,angle=0]{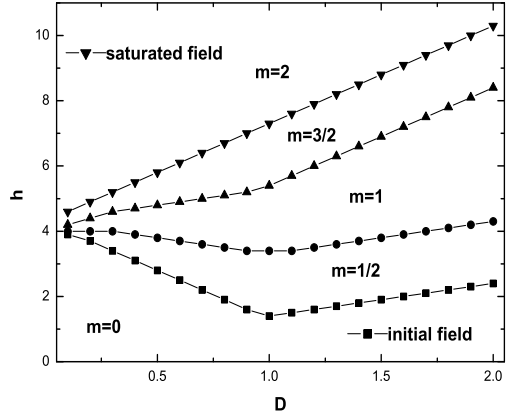}
\caption{\label{fig:Fig12} The magnetization phase diagram of the
ground state of AF spin-$2$ Ising chain under finite magnetic
field. The square-dot line, the circular-dot line, the upper
triangle-dot line, and the lower triangle-dot line represent the
initial critical field, the second critical field, the third
critical field, and the saturated field, respectively.}
\end{center}
\end{figure}
% figure 7
\begin{figure}
\begin{center}
\subfigure[\hspace{0.5cm}]{\label{fig:sub:a}
\includegraphics[width=6cm,height=5.5cm,angle=0]{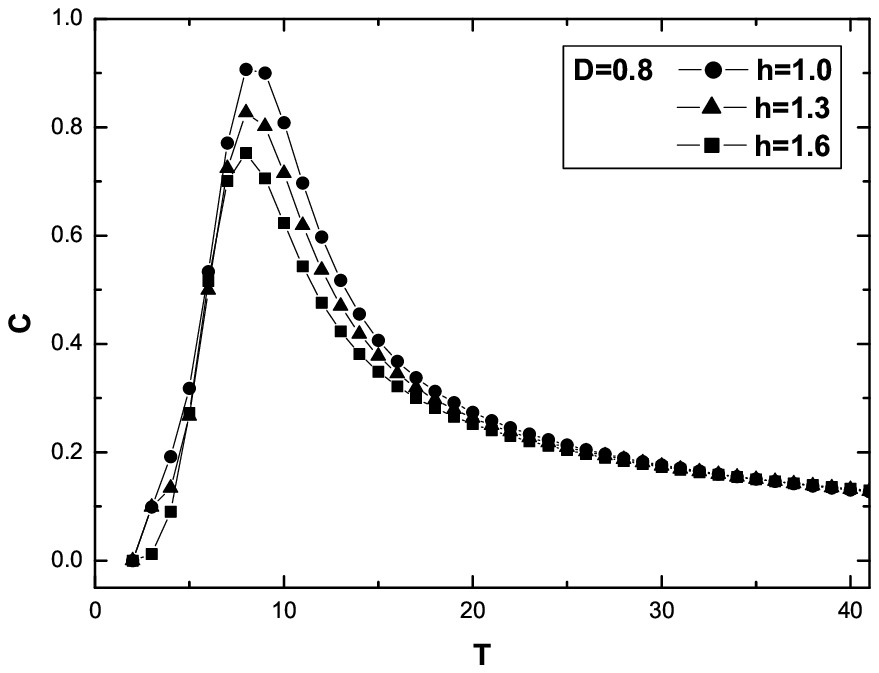}}
\hspace{0.4cm} \subfigure[\hspace{0.5cm}]{\label{fig:sub:b}
\includegraphics[width=6cm,height=5.5cm,angle=0]{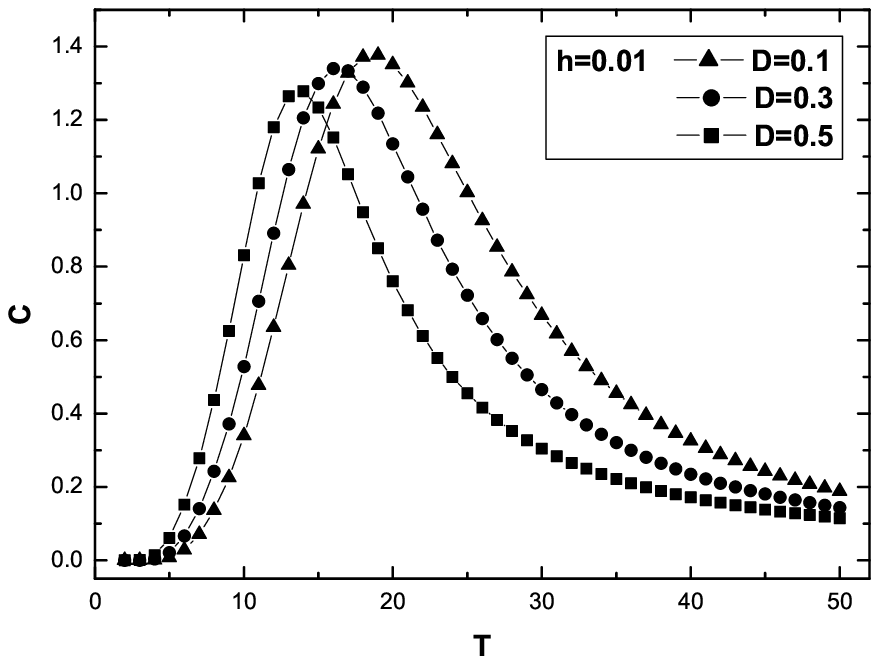}}
\caption{The specific heat $C$ as a function of temperature T
(unit by $J$): (a) $h=1.0, 1.3, 1.6$ at fixed value of $D=0.8$ (b)
$D=0.1, 0.3, 0.5$ at fixed value of $h=0.01$.}
\label{fig:sub:Fig7a-7b}
\end{center}
\end{figure}
% figure 8
\begin{figure}
\begin{center}
\subfigure[\hspace{0.5cm}]{\label{fig:sub:a}
\includegraphics[width=6cm,height=5.5cm,angle=0]{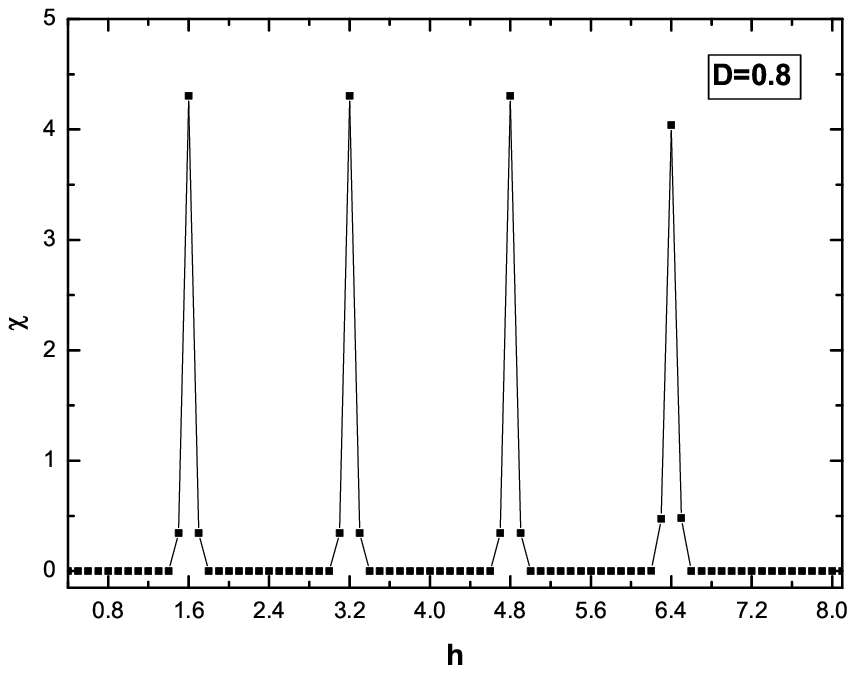}}
\hspace{0.4cm} \subfigure[\hspace{0.5cm}]{\label{fig:sub:b}
\includegraphics[width=6cm,height=5.5cm,angle=0]{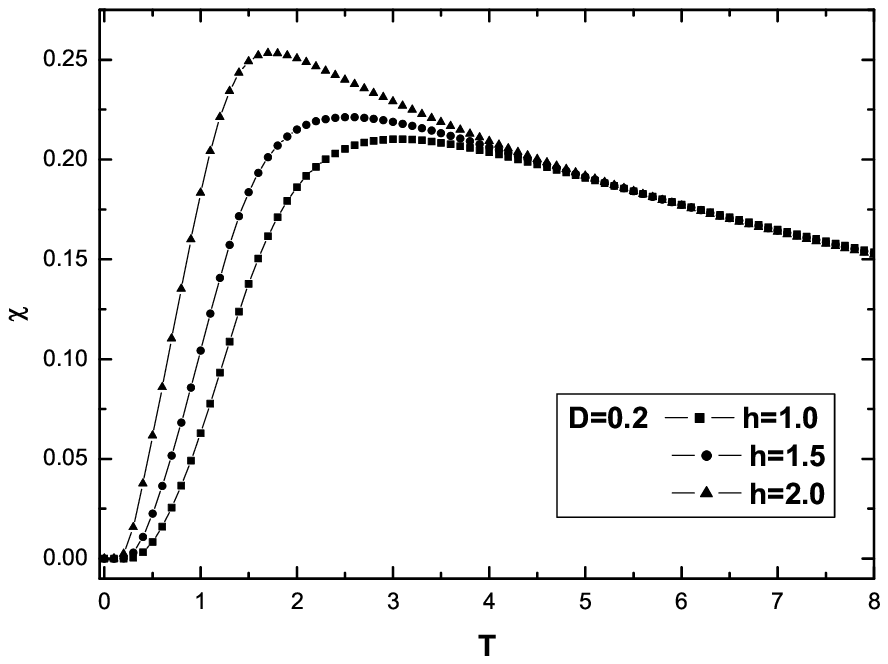}}
\caption{(a) Susceptibility as a function of magnetic field $h$ at
$D=0.8$ and $T=0.01$ (b) Susceptibility as a function of
temperature for various value of $h$ at $D=0.2$ (unit by $J$).}
\label{fig:sub:Fig8a-8b}
\end{center}
\end{figure}

The temperature dependence of the specific heat for different
field values and fixed value of $D$ is shown in Fig.\,7(a) and for
different single ion-anisotropy values and fixed field value of
$h$ is shown in Fig.\,7(b), respectively. One can suggested that
the physical behavior of specific heat of spin-2 chain has the
same characteristic with spin-3/2. In fact, as can be seen from
Fig.\,7(a), the height of the specific heat peak for the fixed
value of $D$ decreases and moves towards zero temperature when the
external field $h$ increases. On the other hand, the height of the
specific heat peak for the fixed value of $h$ decreases and moves
towards zero temperature when the value of single-ion anisotropy
increases in Fig.\,7(b).

Finally, we have plotted the magnetic susceptibility versus
external field for fixed $D$ in Fig.\,8(a) and versus the
temperature for different values of $h$ and fixed value of $D$ in
Fig.\,8(b), respectively. The peaks in Fig.\,8(a) indicate the
critical field values transition from one plateau to another one
for the spin-2 chain. On the other hand, the behavior of the
susceptibility curves in Fig.\,8(b) clearly provide that 1D AF
spin-2 Ising chain has an energy gap which leads to the magnetic
plateaus.

\section{Conclusion}

In this study, using by quasi-classical Hamiltonian(1), we have
evaluated the magnetic and thermal properties of 1D AF spin-$3/2$
and spin-$2$ Ising chain with a single-ion anisotropy in the
presence of an external magnetic field at low temperature in the
framework of the transfer matrix formalism.

The most important result stemming from this study is the
confirmation of a multi step magnetization process by a numerical
calculation for both in the case of a half-odd integer, and an
integer spin quantum $S$. Indeed, we found out that the $2S+1$
step-like magnetization plateaus of the system appear for all
values of positive single-ion anisotropy $D>0$ in both spin
system, when the external field varied from zero to the saturated
field. In addition, the magnetic plateau does not appear for
$D=0.0$ in these spin systems, however, at extreme value of
$J=0.0$ we observed that the spin-$3/2$ Ising chain has two
plateaus at $m=0.5$ and $1.5$ and then spin-$2$ Ising chain has
three plateaus at $m=0.0$, $1.0$, and $2.0$. These evidences
confirm that single ion-anisotropy and the nearest-neighbor
interaction contribute to be formed the magnetic plateaus.
Therefore, plateau forms in the 1D AF spin-$S$ Ising chains can be
explain owing to the co-existence of antiferromagnetic interaction
of $(S^{z}_{i}S^{z}_{i+1})$ and positive single-ion anisotropy
$(S^{z}_{i})^2$.

The plateau predictions in this study are compatible with previous
theoretical studies
\cite{oshikawa,chen,sakaitakahashi,aydiner1,ohanyan,aydiner2}.
Therefore, the present results provide that there is a class of
one-dimensional spin systems which in terms of Heisenberg and
Ising spins lead to the qualitatively same structures of the
magnetization profiles, particularly to the formation of
magnetization plateaus. Hence, we suggest that the quasi-classical
approach can be used to examine the magnetic and thermal behaviors
of 1D AF spin-$S$ Ising chains as well as the Heisenberg
Hamiltonian.

Unfortunately, we have not compare our thermodynamical results
with the experimental results since there are no enough
experimental studies for these systems in the literature. However,
it is understand from present theoretical evidences that the
magnetic and thermal characteristic of the 1D AF spin-$S$ Ising
chains are the same. Therefore, we expected that our results will
be agreement with the experimental results of the ideal spin-$3/2$
and spin-$2$ compounds.

Finally, we stated that one does not expect any spin-Peierls
transition in these models, because there is no physical mechanism
for it within a model constructed purely from $S^z$ operator,
without any coupling to the lattice. A spin-Peierls transition may
occurs only due to a small amount of magnetic impurities and
derivations from stoichiometry of the compound synthesized
experimentally \cite{vasilier}.

\end{document}